\documentclass[11pt]{article}
\usepackage{latexsym,amsmath,amsfonts,graphics}
\begin{document}
\begin{center}
{\LARGE\textbf{Leader-following Consensus Problems with a Time-varying Leader under Measurement Noises}}\\
\bigskip
\bigskip
Yilun Shang\footnote{email: \texttt{shyl@sjtu.edu.cn}}\bigskip

\footnotesize Department of Mathematics, Shanghai Jiao Tong
University, Shanghai 200240, P. R. China.

\normalsize
\end{center}

\begin{abstract}
In this paper, we consider a leader-following consensus problem for
networks of continuous-time integrator agents with a time-varying
leader under measurement noises. We propose a neighbor-based
state-estimation protocol for every agent to track the leader, and
time-varying consensus gains are introduced to attenuate the noises.
By combining the tools of stochastic analysis and algebraic graph
theory, we study mean square convergence of this multi-agent system
under directed fixed as well as switching interconnection
topologies. Sufficient conditions are given for mean square
consensus in both cases. Finally, a numerical example is given to
illustrate our theoretical results.

\end{abstract}

\bigskip

\normalsize

\noindent{\Large\textbf{1. Introduction}}
\smallskip

In recent years, there has been an increasing research interest in
the distributed coordination for multi-agent systems. This is partly
due to its broad applications in many areas such as cooperative
control of unmanned aerial vehicles, formation control
\cite{13,14,99} and swarming behaviors of social living beings
\cite{27,15,16}.

Consensus problems have a long history in computer science and
formed the foundation of the field of distributed computing
\cite{17}. In consensus control, it is critical to design a
decentralized network algorithm based on neighborhood information
for agents to reach an agreement on their states, asymptotically in
some sense. For a variety of consensus algorithms and convergence
results we refer the reader to the comprehensive surveys
\cite{11,12} and references therein. Most researches in the previous
literature assume the exchange of messages between agents is
error-free. However, this is only an ideal approximation for real
communication processes. Recently, consensus of dynamic networks
with random measurement noises has attracted the attention of some
researchers. In \cite{19,18}, the authors introduce time-varying
consensus gains and design control schemes based on a Kalman filter
structure. The decreasing consensus gain $a(k)$ (where $k$ is the
discrete time instant) in the protocols is proposed in \cite{2} to
attenuate the measurement noises in a strongly connected circulant
network. The analysis in \cite{2} is generalized to strongly
connected digraphs in \cite{10} and digraphs containing a spanning
tree in \cite{20} by the same authors. The work in \cite{5} deals
with discrete-time average consensus problems in switching balanced
digraphs under stochastic communication noises, while \cite{1}
investigates the continuous-time average consensus control with
fixed topology and Gaussian communication noises. The authors in
\cite{6} treat a continuous-time leader-following consensus control
under measurement noises with a constant state leader.

In this paper, motivated by the above works, we consider a
leader-following consensus problem for networks of continuous-time
integrator agents with a time-varying leader in directed fixed and
switching topologies. The control input of each agent is based on
the measurement of its neighbors' states and some estimated data of
the leader which are both corrupted by white noises. We design a
leader-following consensus protocol such that the leader has an
underlying dynamics and some variables (e.g. velocity and
acceleration) of the leader cannot be measured and every follower
can obtain the measured information (e.g. position) of the leader
only when they are connected with the leader directly. The
collective behavior of self-organized groups of agents with active
(or dynamical) leaders is one of the most interesting topics in
distributed cooperative control. However, as \cite{32} suggests, the
extension of consensus algorithms from a constant reference to a
time-varying one is non-trivial. Some related results can be found
e.g. in \cite{7,8,21}, where the systems considered are all
deterministic and free of noise.

Inspired by \cite{2,1,6}, we introduce time-varying consensus gains
in the followers control protocol to attenuate the measurement
noises, which lead to a time-varying stochastic differential
equation of the system. The state matrix of the equation is
time-dependent and no longer a Laplacian matrix, and is neither
symmetric nor diagonalizable. To implement the convergence study, we
merge stochastic analysis and algebraic graph theory, by developing
a Lyapunov-based approach and addressing the It\^o integral by the
stopping time truncation method. Firstly, we derive a sufficient
condition for the state of each follower to converge to that of the
leader in mean square under fixed and directed interconnection
topology. Then it is shown that the algorithm also render each
follower track the leader in mean square under switching topology
when the subgraph induced by the followers is balanced.

The rest of the paper is organized as follows. In Section 2, we
provide some preliminaries and present the leader-following
consensus protocol. Section 3 contains the convergence analyses
under directed fixed and switching interaction topologies. A
numerical example is given in Section 4 and we conclude the paper in
Section 5.

\bigskip
\noindent{\Large\textbf{2. Problem formulation}}
\smallskip

Before we proceed, some basic concepts on graph theory (see e.g.
\cite{23}) are provided as below.

Let $\mathcal{G}=(\mathcal{V},\mathcal{E},\mathcal{A})$ be a
weighted digraph with the set of vertices
$\mathcal{V}=\{1,2,\cdots,n\}$ and the set of arcs
$\mathcal{E}\subseteq\mathcal{V}\times\mathcal{V}$. The vertex $i$
in $\mathcal{G}$ represents the $i$th agent, and a directed edge
$(i,j)\in\mathcal{E}$ means that agent $j$ can directly receive
information from agent $i$. The set of neighbors of vertex $i$ is
denoted by $\mathcal{N}_i=\{j\in\mathcal{V}|\
(j,i)\in\mathcal{E}\}$. $\mathcal{A}=(a_{ij})\in\mathbb{R}^{n\times
n}$ is called the weighted adjacency matrix of $\mathcal{G}$ with
nonnegative elements and $a_{ij}>0$ if and only if
$j\in\mathcal{N}_i$. The in-degree and out-degree of vertex $i$ are
defined as $d_{in}(i)=\sum_{j=1}^na_{ij}$ and
$d_{out}(i)=\sum_{j=1}^na_{ji}$, respectively. If
$d_{in}(i)=d_{out}(i)$ for $i=1,2,\cdots,n$, then the digraph
$\mathcal{G}$ is called balanced \cite{22}. The Laplacian of
$\mathcal{G}$ is defined as $L=D-\mathcal{A}$, where
$D=\mathrm{diag}(d_{in}(1), d_{in}(2),\cdots, d_{in}(n))$. A digraph
$\mathcal{G}$ is called strongly connected if there is a directed
path from $i$ to $j$ between any two distinct vertices
$i,j\in\mathcal{V}$. There exists a directed path from vertex $i$ to
vertex $j$, then $j$ is said to be reachable from $i$. For every
vertex in digraph $\mathcal{G}$, if there is a path from vertex $i$
to it, then we say $i$ is globally reachable in $\mathcal{G}$. This
is much weaker than strong connectedness.

Here, we consider a system consisting of $n+1$ agents, in which an
agent indexed by $0$ is assigned as the leader and the other agents
indexed by $1,2,\cdots,n$ are referred as follower agents. The
information interaction topology among $n$ followers are described
by the digraph $\mathcal{G}$ as defined above; and the whole system
including $n+1$ agents is conveniently modeled by a weighted digraph
$\overline{\mathcal{G}}=(\overline{\mathcal{V}},\overline{\mathcal{E}},\overline{\mathcal{A}})$
with $\overline{\mathcal{V}}=\{0,1,\cdots,n\}$ and
$$
\overline{\mathcal{A}}=\left(\begin{array}{cccc}
0&0&\cdots&0\\
a_{10}&a_{11}&\cdots&a_{1n}\\
\vdots&\vdots&\ddots&\vdots\\
a_{n0}&a_{n1}&\cdots&a_{nn}\\
\end{array}\right)\in\mathbb{R}^{(n+1)\times(n+1)},
$$
where the lower right block submatrix of order $n$ can be viewed as
$\mathcal{A}$. We define a diagonal matrix
$B=\mathrm{diag}(b_1,b_2,\cdots,b_n)$ to be the leader adjacency
matrix associated with $\overline{\mathcal{G}}$, where
$b_i=a_{i0}\ge0$ and $b_i>0$ if and only if
$0\in\mathcal{N}_i(\overline{\mathcal{G}})$. Here,
$\mathcal{N}_i(\overline{\mathcal{G}})$ is the set of neighbors of
agent $i$ in $\overline{\mathcal{G}}$.

The continuous-time dynamics of $n$ followers is described as
follows:
\begin{equation}
\dot{x}_i(t)=u_i(t),\quad i=1,2,\cdots,n,\label{1}
\end{equation}
where $x_i(t)\in\mathbb{R}$ is the state of the $i$th agent, and
$u_i(t)\in\mathbb{R}$ is the control input. The leader of this
considered multi-agent system is described by a double integrator of
the form:
\begin{equation}
\left\{\begin{array}{c} \dot{x}_0(t)=g(t)v_0(t)\\
\dot{v}_0(t)=a_0(t)\\
y(t)=x_0(t)\label{2}
\end{array}
\right.
\end{equation}
where $g(t):[0,\infty)\rightarrow(0,\infty)$ is piecewise
continuous, $y(t)$ is the measured output and $a_0(t)$ is the input.
We assume $g(t)$ and $a_0(t)$ are known, that is, the dynamical
behavior of the leader is precisely known (c.f. Remark 1). On the
other hand, $y(t)=x_0(t)$ is the only data that may be gotten by the
followers when they are connected to the leader directly. Since
$v_0(t)$ cannot be measured, we have to estimate $v_0(t)$ in a
distributed way during the evolution. The estimate of $v_0(t)$ by
agent $i$ is denoted by $v_i(t)$, $i=1,2,\cdots,n$.

In our model, the $i$th agent receives information from its
neighbors with measurement noises:
\begin{equation}
y_{ji}(t)=x_j(t)+\sigma_{ji}n_{ji}(t), \quad
j\in\mathcal{N}_i\label{3},
\end{equation}
\begin{equation}
y_{0i}(t)=x_0(t)+\sigma_{0i}n_{0i}(t),\label{4}
\end{equation}
where $y_{ji}(t)$ $(i\in\mathcal{V}, j\in\overline{\mathcal{V}})$
denotes the measurement of the $j$th agent's state $x_j(t)$ by the
$i$th agent. The $\{n_{ji}(t)|\ j\in\overline{\mathcal{V}},
i\in\mathcal{V}\}$ are independent standard white noises and
$\sigma_{ji}\ge0$ is the noise intensity.

A group of controls $\mathcal{U}=\{u_i|\ i=1,2,\cdots,n\}$ is called
a measurement-based distributed protocol \cite{1}, if
$u_i(t)\in\sigma\big(x_i(s),\bigcup_{{j\in\mathcal{N}}_i}y_{ji}(s),\
0\le s\le t\big)$, for $t\ge0$, $i=1,2,\cdots,n$. Herein
$\sigma(\xi_{\lambda},\lambda\in\Lambda)$ denotes the
$\sigma$-algebra generated by a family of random variables
$\{\xi_{\lambda},\lambda\in\Lambda\}$. The so-called
leader-following consensus problem is to design a measurement-based
distributed protocol such that each follower's state will converge
to the leader's in some sense as time goes on.

Consequently, we propose the distributed control protocol which
consists of two parts: \\
$\bullet$ a neighbor-based feedback law:
\begin{equation}
u_i(t)=h(t)\Big(\sum_{j\in\mathcal{N}_i}a_{ij}(y_{ji}(t)-x_i(t))+b_i(y_{0i}(t)-x_i(t))\Big)+g(t)v_i(t),\quad
i=1,2,\cdots,n\label{5}
\end{equation}
where $t\ge0$ and $h(t):[0,\infty)\rightarrow(0,\infty)$ is a
piecewise
continuous function, called a time-varying consensus gain \cite{1}.\\
$\bullet$ a dynamic neighbor-based system to estimate $v_0(t)$:
\begin{equation}
\dot{v}_i(t)=a_0(t)+\gamma
h(t)\Big(\sum_{j\in\mathcal{N}_i}a_{ij}(y'_{ji}(t)-x_i(t))+b_i(y'_{0i}(t)-x_i(t))\Big),\quad
i=1,2,\cdots,n\label{6}
\end{equation}
where $0<\gamma<1$ is some constant, and moreover $y'_{ji}(t)$ and
$y'_{0i}(t)$ are independent copies of $y_{ji}(t)$ and $y_{0i}(t)$,
respectively. In other words, we have
\begin{equation}
y'_{ji}(t)=x_j(t)+\sigma_{ji}n'_{ji}(t), \quad
j\in\mathcal{N}_i\label{7},
\end{equation}
\begin{equation}
y'_{0i}(t)=x_0(t)+\sigma_{0i}n'_{0i}(t),\label{8}
\end{equation}
where $\{n'_{ji}(t)|\ j\in\overline{\mathcal{V}}, i\in\mathcal{V}\}$
are independent standard white noises and independent with
$\{n_{ji}(t)|\ j\in\overline{\mathcal{V}}, i\in\mathcal{V}\}$.

The set of neighbors $\mathcal{N}_i$ of agent $i$ in (\ref{5}) and
(\ref{6}) varies when the interconnection topology is switching and
we defer the corresponding protocol formulation to Section 3.2.

\noindent\textbf{Remark 1.}\itshape \quad We take individual state
$x_i$ as scalar for simplicity in (\ref{1}) and it can be extended
to multi-dimensional scenarios as studied in \cite{8,21} without
much effort. For example, if $x_i\in\mathbb{R}^2$, it can be thought
as the position of agent $i$ moving in a plane. Therefore, $gv_0$
and $ga_0+\dot{g}v_0$ are the velocity and acceleration of the
leader respectively, which are known since the exact dynamics of the
leader is assumed. \normalfont

\noindent\textbf{Remark 2.}\itshape \quad We separate a factor $g$
from the `velocity term' of the leader in (\ref{2}) in order to tone
the decreasing consensus gain h, which appears to be a notable
feature distinct from some kinds of uncertain environment (see e.g.
\cite{24,25}), where a random term is directly appended behind the
equation of the system. In such works, the consensus gains are
supposed to have positive lower bound. \normalfont

\noindent\textbf{Remark 3.}\itshape \quad From (\ref{5}) and
(\ref{6}) it is clear that the designed protocol for the $i$th agent
is indeed a measurement-based distributed protocol since it relies
only on the state of itself and its neighbors. \normalfont

Let $x(t)=(x_1(t),\cdots,x_n(t))^T$ and
$v(t)=(v_1(t),\cdots,v_n(t))^T$. Denote the $i$th row of the matrix
$\mathcal{A}$ by $\alpha_i$, and
$\Sigma_i:=\mathrm{diag}(\sigma_{1i},\cdots,\sigma_{ni})$ for
$i=1,2,\cdots,n$. Hence
$\Sigma:=\mathrm{diag}(\alpha_1\Sigma_1,\cdots,\alpha_n\Sigma_n)$ is
an $n\times n^2$ dimensional block diagonal matrix. Let
$n_0(t)=(n_{01}(t),\cdots,n_{0n}(t))^T$ and
$n_i(t)=(n_{1i}(t),\cdots,n_{ni}(t))^T$ for $i=1,2,\cdots,n$. In
addition, $n'_0(t)$ and $n'_i(t)$ can be defined in a similar way.
The juxtaposed matrix $Q:=(B,\Sigma)$ is an $n\times n(n+1)$
dimensional block matrix. Combining (\ref{1}) with (\ref{5}) and
(\ref{6}), we may write the protocol in a matrix form:
\begin{equation}
\left\{\begin{array}{rl}
\frac{\mathrm{d}x(t)}{\mathrm{d}t}=&-h(t)(L+B)x(t)+h(t)B1x_0(t)+g(t)v(t)+h(t)QZ(t)\\
\frac{\mathrm{d}v(t)}{\mathrm{d}t}=&a_0(t)1-\gamma
h(t)(L+B)x(t)+\gamma h(t)B1x_0(t)+\gamma h(t)QZ'(t)\label{9}
\end{array}\right.
\end{equation}
where $Z(t)=(n_0^T(t),n_1^T(t),\cdots,n_n^T(t))^T$ and
$Z'(t)=(n'_0{}^T(t),n'_1{}^T(t),\cdots,n'_n{}^T(t))^T$ are two
$n(n+1)$ dimensional independent standard white noise sequences, and
$1=(1,\cdots,1)^T\in\mathbb{R}^n$. The system (\ref{9}) may be
further written in the form of the It\^o stochastic differential
equations:
\begin{equation}
\left\{\begin{array}{cl}
\mathrm{d}x(t)=&-h(t)(L+B)x(t)\mathrm{d}t+h(t)B1x_0(t)\mathrm{d}t+g(t)v(t)\mathrm{d}t+h(t)G\mathrm{d}W_1(t)\\
\mathrm{d}v(t)=&a_0(t)1\mathrm{d}t-\gamma
h(t)(L+B)x(t)\mathrm{d}t+\gamma h(t)B1x_0(t)\mathrm{d}t+\gamma
h(t)G\mathrm{d}W_2(t)\label{10}
\end{array}\right.
\end{equation}
where $W_1(t)=(W_{11}(t),\cdots,W_{1n}(t))^T$ and
$W_2(t)=(W_{21}(t),\cdots,W_{2n}(t))^T$ are two $n$ dimensional
standard Brownian motions which are independent with each other, and
$G:=\mathrm{diag}\Big(\sqrt{b_1^2+\sum_{j\in\mathcal{N}_1}\sigma_{j1}^2a_{1j}^2}\
,\cdots,\sqrt{b_n^2+\sum_{j\in\mathcal{N}_n}\sigma_{jn}^2a_{nj}^2}\
\Big)$.

\bigskip
\noindent{\Large\textbf{3. Convergence analysis}}
\smallskip

In this section we will give the convergence analysis of the system
(\ref{10}) and show that the state of every follower will track that
of the leader in the sense of mean square convergence, that is,
$E\|x(t)-x_0(t)1\|\rightarrow0$, as $t\rightarrow\infty$. Here
$\|\cdot\|$ denotes Frobenius norm. Two different cases, fixed
topology and switching topology, are explored.

\noindent\textbf{Remark 4.}\itshape \quad Mean square consensus
protocols for stochastic systems are first introduced in \cite{2}
and then further studied by several researchers (e.g.
\cite{20,10,1,5,6}). Mean square convergence seems to be an
important alternative for almost sure convergence in consensus
problems under noisy environments. \normalfont

For a given symmetric matrix $A$, let $\lambda_{\max}(A)$ and
$\lambda_{\min}(A)$ denote its maximum and minimum eigenvalue,
respectively. To get the main result, we need the following
assumptions:\\
\textbf{(A1)}\quad The vertex 0 is globally reachable in $\overline{\mathcal{G}}$.\\
\textbf{(A2)}\quad There is a $\delta>0$, such that
$\frac{h(t)}{g(t)}>\frac{\lambda_{\max}(P)}{2\gamma(1-\gamma^2)}+\delta$ for $t\ge 0$.\\
Here, $P$ is a positive definite matrix defined by Equation (\ref{13}), see below.\\
\textbf{(A3)}\quad $\int_0^{\infty}h(s)\mathrm{d}s=\infty$.\\
\textbf{(A4)}\quad $\int_0^{\infty}h^2(s)\mathrm{d}s<\infty$.

\noindent\textbf{Remark 5.}\itshape \quad Assumption (A1) is imposed
on the network topology, which is much weaker than strong
connectedness. The technical Assumption (A2) roughly means that $g$
is comparable with the consensus gain $h$. Assumptions (A3) and (A4)
are called convergence condition and robustness condition
respectively in \cite{1}, and which are standard assumptions often
used in the stochastic approximation \cite{28}. \normalfont

\bigskip
\noindent{\large\textbf{3.1. Fixed topology}}
\smallskip

Let $x^*=x-x_01$ and $v^*=v-v_01$. We then obtain an error dynamics
of (\ref{10}) as follows:
\begin{equation}
\mathrm{d}\varepsilon(t)=F(t)\varepsilon(t)\mathrm{d}t+G(t)\mathrm{d}W(t),\quad
t\ge 0\label{11}
\end{equation}
where $\varepsilon(t)=(x^*(t),v^*(t))^T$, $W(t)=(W_1(t),W_2(t))^T$
and
$$
F(t)=\left(\begin{array}{cc}
-h(t)(L+B)&g(t)I_n\\-\gamma h(t)(L+B)&0
\end{array}\right),\qquad
G(t)=h(t)\left(\begin{array}{c} G\\ \gamma G
\end{array}\right):=h(t)\widetilde{G}.
$$
Here $I_n$ denotes the $n\times n$ dimensional identity matrix.

We will need a lemma for Laplacian matrix.

\smallskip
\noindent\textbf{Lemma 1.}\cite{23,29}\itshape \quad The Laplacian
matrix $L$ of a digraph
$\mathcal{G}=(\mathcal{V},\mathcal{E},\mathcal{A})$ has at least one
zero eigenvalue and all of the nonzero eigenvalues are in the open
right half plane. Furthermore, $L$ has exactly one zero eigenvalue
if and only if there is a globally reachable vertex in
$\mathcal{G}$.\normalfont

The main result in this section is given as follows:

\smallskip
\noindent\textbf{Theorem 1.}\itshape \quad For system (\ref{1}) with
the consensus protocols (\ref{5}) and (\ref{6}), if Assumptions
(A1)-(A4) hold, then
\begin{equation}
\lim_{t\rightarrow\infty}E\|\varepsilon(t)\|^2=0.\label{12}
\end{equation}
 \normalfont

\smallskip
\noindent\textbf{Proof.} By Assumption (A1) and Lemma 1, we know
$L+B$ is a positive stable matrix, or in other words, $-L-B$ is a
stable matrix. From Lyapunov theorem, there is a unique positive
definite matrix $P\in\mathbb{R}^{n\times n}$ such that
\begin{equation}
(L+B)^TP+P(L+B)=I_n.\label{13}
\end{equation}
Let $\widetilde{P}=\left(\begin{array}{cc}P&-\gamma P\\-\gamma P&P
\end{array}\right)$ and define a Lyapunov function
$V(t)=\varepsilon^T(t)\widetilde{P}\varepsilon(t)$. Since
$0<\gamma<1$, $\widetilde{P}$ is a positive definite matrix. In
fact, we have
$\lambda_{\min}(\widetilde{P})=(1-\gamma)\lambda_{\min}(P)$ and
$\lambda_{\max}(\widetilde{P})=(1+\gamma)\lambda_{\max}(P)$.
Utilizing It\^o formula and (\ref{11}), we have
$$
\mathrm{d}V(t)=\varepsilon^T(t)\big(\widetilde{P}F(t)+F^T(t)\widetilde{P}\big)\varepsilon(t)\mathrm{d}t+\mathrm{tr}(G^T(t)\widetilde{P}G(t))\mathrm{d}t+2\varepsilon^T(t)\widetilde{P}G(t)\mathrm{d}W(t).
$$
Here $\mathrm{tr}(\cdot)$ means the trace of a matrix. From the
Lyapunov equation (\ref{13}), we get
$$
\widetilde{P}F(t)+F^T(t)\widetilde{P}=-h(t)\left(\begin{array}{cc}(1-\gamma^2)I_n&-P\frac{g(t)}{h(t)}\\
-P\frac{g(t)}{h(t)}&2\gamma P\frac{g(t)}{h(t)}
\end{array}\right):=-h(t)\widetilde{Q}(t).
$$
Invoking the Haynsworth inertia additivity formula \cite{30},
Assumption (A2) and the positive definiteness of $P$, we know that
$\widetilde{Q}(t)$ is positive definite with
$\rho:=\min_{t\ge0}\big\{\lambda_{\min}(\widetilde{Q}(t))\big\}>0$.

Thereby, we have
\begin{equation}
\mathrm{d}V(t)\le-h(t)\frac{\rho}{\lambda_{\max}(\widetilde{P})}V(t)\mathrm{d}t+h^2(t)\mathrm{tr}(\widetilde{G}^T\widetilde{P}\widetilde{G})\mathrm{d}t+2h(t)\varepsilon^T(t)\widetilde{P}\widetilde{G}\mathrm{d}W(t).\label{14}
\end{equation}

Next we want to prove
\begin{equation}
E\int_{t_0}^th(s)\varepsilon^T(s)\widetilde{P}\widetilde{G}\mathrm{d}W(s)=0,
\quad \forall\ 0\le t_0\le t.\label{15}
\end{equation}
For any given $T\ge t_0\ge0$ and $K\in\mathbb{N}$, let
$\tau_K^{t_0,T}=T\wedge\inf\{t\ge t_0|\
\varepsilon^T(t)\widetilde{P}\varepsilon(t)\ge K\}$, which is a
stopping time. By (\ref{14}) we have, for $t_0\le t\le T$,
\begin{eqnarray*}
\lefteqn{E\big(V(t\wedge\tau_K^{t_0,T})1_{[t\le\tau_K^{t_0,T}]}\big)-E
V(t_0)}\\
 &&\le-\frac{\rho}{\lambda_{\max}(\widetilde{P})}\int_{t_0}^th(s)E\big(V(s\wedge\tau_K^{t_0,T})1_{[s\le\tau_K^{t_0,T}]}\big)\mathrm{d}s+\mathrm{tr}(\widetilde{G}^T\widetilde{P}\widetilde{G})\int_{t_0}^th^2(s)\mathrm{d}s\\
&&\le\mathrm{tr}(\widetilde{G}^T\widetilde{P}\widetilde{G})\int_{t_0}^Th^2(s)\mathrm{d}s.\label{16}
\end{eqnarray*}
This implies that there is a constant $C_{t_0,T}>0$ such that
$$
E\big(V(t\wedge\tau_K^{t_0,T})1_{[t\le\tau_K^{t_0,T}]}\big)\le
C_{t_0,T},\quad \forall\  t_0\le t\le T.
$$
Since $\lim_{K\rightarrow\infty}t\wedge\tau_K^{t_0,T}=t$ a.s., for
$t_0\le t\le T$, by Fatou lemma, we derive
$$
\sup_{t_0\le t\le T}EV(t)\le C_{t_0,T}.
$$
Accordingly,
$$
E\int_{t_0}^th^2(s)V(s)\mathrm{d}s\le\sup_{t_0\le s\le
t}EV(s)\cdot\int_0^Th^2(s)\mathrm{d}s<\infty,\quad\forall\ 0\le
t_0\le t.
$$
Combining this with
$$
E\int_{t_0}^th^2(s)\|\varepsilon^T(s)\widetilde{P}\widetilde{G}\|^2\mathrm{d}s\le\|\widetilde{P}\|\|\widetilde{G}\|^2E\int_{t_0}^th^2(s)V(s)\mathrm{d}s
$$
yields (\ref{15}) (c.f. \cite{4}).

Now employing (\ref{14}) and (\ref{15}), we obtain
$$
EV(t)-EV(0)\le-\frac{\rho}{\lambda_{\max}(\widetilde{P})}\int_0^th(s)EV(s)\mathrm{d}s+\mathrm{tr}(\widetilde{G}^T\widetilde{P}\widetilde{G})\int_0^th^2(s)\mathrm{d}s,\quad
\forall\ t\ge0.
$$
Thus, from the comparison principle \cite{3},
\begin{eqnarray}
EV(t)&\le&
EV(0)\exp\Big(-\frac{\rho}{\lambda_{\max}(\widetilde{P})}\int_0^th(s)\mathrm{d}s\Big)\nonumber\\
& &+\
\mathrm{tr}(\widetilde{G}^T\widetilde{P}\widetilde{G})\int_0^th^2(s)\exp\Big(-\frac{\rho}{\lambda_{\max}(\widetilde{P})}\int_s^th(u)\mathrm{d}u\Big)\mathrm{d}s.\label{16}
\end{eqnarray}
Clearly, by Assumption (A3) the first term on the right hand side of
(\ref{16}) tends to 0, as $t\rightarrow\infty$. For any $\eta>0$, by
Assumption (A4), there exists some $s_0>0$ such that
$\int_{s_0}^{\infty}h^2(s)\mathrm{d}s<\eta$. Hence,
\begin{eqnarray*}
\lefteqn{\int_0^th^2(s)\exp\Big(-\frac{\rho}{\lambda_{\max}(\widetilde{P})}\int_s^th(u)\mathrm{d}u\Big)\mathrm{d}s}\\
&
&=\int_0^{s_0}h^2(s)\exp\Big(-\frac{\rho}{\lambda_{\max}(\widetilde{P})}\int_s^th(u)\mathrm{d}u\Big)\mathrm{d}s+\int_{s_0}^th^2(s)\exp\Big(-\frac{\rho}{\lambda_{\max}(\widetilde{P})}\int_s^th(u)\mathrm{d}u\Big)\mathrm{d}s\\
&
&\le\exp\Big(-\frac{\rho}{\lambda_{\max}(\widetilde{P})}\int_{s_0}^th(u)\mathrm{d}u\Big)\int_0^{s_0}h^2(s)\mathrm{d}s+\int_{s_0}^th^2(s)\mathrm{d}s\\
&
&\le\exp\Big(-\frac{\rho}{\lambda_{\max}(\widetilde{P})}\int_{s_0}^th(u)\mathrm{d}u\Big)\int_0^{\infty}h^2(s)\mathrm{d}s+\eta,\quad
\forall\  t\ge s_0
\end{eqnarray*}
By Assumptions (A3), (A4) and the arbitrariness of $\eta$, the last
expression tends to zero, as $t\rightarrow\infty$. Therefore,
(\ref{16}) yields $\lim_{t\rightarrow\infty}EV(t)=0$. Note that
$$
\|\varepsilon(t)\|^2\le\frac{V(t)}{\lambda_{\min}(\widetilde{P})}
$$
which concludes the proof. $\Box$

\noindent\textbf{Remark 6.}\itshape \quad As is known, the solution
to Lyapunov matrix equation may be obtained by using Kronecker
product. Thus the positive definite matrix $P$ involved in
Assumption (A2) can be given explicitly. \normalfont

\smallskip
\noindent\textbf{Remark 7.}\itshape \quad Theorem 1 implies that in
the fixed topology, under Assumptions (A1)-(A4), the designed
protocol guarantees the state of each follower tracks that of the
leader in mean square. \normalfont

\bigskip
\noindent{\large\textbf{3.2. Switching topology}}
\smallskip

In this section we deal with the convergence of the protocol under
switching topology.

Let
$\sigma(t):[0,\infty)\rightarrow\mathcal{S}_{\mathcal{H}}=\{1,2,\cdots,N\}$
be a switching signal that determines the communication topology.
The set $\mathcal{H}$ is a set of digraphs with the common vertex
set $\overline{\mathcal{V}}$ and can be denoted as
$\mathcal{H}=\{\overline{\mathcal{G}}_1,\overline{\mathcal{G}}_2,\cdots,\overline{\mathcal{G}}_N\}$,
where $N$ is the total number of digraphs in $\mathcal{H}$.
Naturally, let $\mathcal{G}_{\sigma(t)}$ be the subgraph of
$\overline{\mathcal{G}}_{\sigma(t)}$ induced by $\mathcal{V}$.
Thereby, we rewrite the consensus protocols (\ref{5}) and (\ref{6})
as:
\begin{multline}
u_i(t)=h(t)\Big(\sum_{j\in\mathcal{N}_i(\sigma(t))}a_{ij}(\sigma(t))(y_{ji}(t)-x_i(t))+b_i(\sigma(t))(y_{0i}(t)-x_i(t))\Big)\\
 +g(t)v_i(t),\quad i=1,2,\cdots,n\label{17}
\end{multline}
and
\begin{multline}
\dot{v}_i(t)=a_0(t)+\gamma
h(t)\\
\cdot\Big(\sum_{j\in\mathcal{N}_i(\sigma(t))}a_{ij}(\sigma(t))(y'_{ji}(t)-x_i(t))+b_i(\sigma(t))(y'_{0i}(t)-x_i(t))\Big),\quad
i=1,2,\cdots,n\label{18}
\end{multline}
where, $\mathcal{N}_i(\sigma(t))$ is the set of neighbors of agent
$i$ in the digraph $\mathcal{G}_{\sigma(t)}$ formed by $n$
followers; $a_{ij}(\sigma(t))$ is the $(i,j)$-th element of the
adjacency matrix of $\mathcal{G}_{\sigma(t)}$, and let
$B_{\sigma(t)}:=\mathrm{diag}\big(b_1(\sigma(t)),b_2(\sigma(t)),\cdots,b_n(\sigma(t))\big)$
represent the leader adjacency matrix associated with
$\overline{\mathcal{G}}_{\sigma(t)}$ such that $b_i(\sigma(t))>0$ if
and only if $0\in\mathcal{N}_i(\overline{\mathcal{G}}_{\sigma(t)})$.

In parallel with Section 2, substituting the protocols (\ref{17}),
(\ref{18}) to the system (\ref{1}), we can describe the system in
the form of the It\^o differential equations:
\begin{equation}
\left\{\begin{array}{cl}
\hspace{-5pt}\mathrm{d}x(t)=&\hspace{-5pt}-h(t)(L_{\sigma(t)}+B_{\sigma(t)})x(t)\mathrm{d}t+h(t)B_{\sigma(t)}1x_0(t)\mathrm{d}t+g(t)v(t)\mathrm{d}t+h(t)G_{\sigma(t)}\mathrm{d}W_1(t)\\
\hspace{-5pt}\mathrm{d}v(t)=&\hspace{-5pt}a_0(t)1\mathrm{d}t-\gamma
h(t)(L_{\sigma(t)}+B_{\sigma(t)})x(t)\mathrm{d}t+\gamma
h(t)B_{\sigma(t)}1x_0(t)\mathrm{d}t+\gamma
h(t)G_{\sigma(t)}\mathrm{d}W_2(t)\label{19}
\end{array}\right.
\end{equation}
where $L_{\sigma(t)}$ is the Laplacian matrix of
$\mathcal{G}_{\sigma(t)}$, and
$$
G_{\sigma(t)}:=\mathrm{diag}\Big(\sqrt{b_1^2(\sigma(t))+\sum_{j\in\mathcal{N}_1(\sigma(t))}\sigma_{j1}^2a_{1j}^2(\sigma(t))}\
,\cdots,\sqrt{b_n^2(\sigma(t))+\sum_{j\in\mathcal{N}_n(\sigma(t))}\sigma_{jn}^2a_{nj}^2(\sigma(t))}\
\Big).
$$

Let $x^*=x-x_01$ and $v^*=v-v_01$ as in Section 3.1. We get an error
dynamics of (\ref{19}) as follows:
\begin{equation}
\mathrm{d}\varepsilon(t)=F_{\sigma}(t)\varepsilon(t)\mathrm{d}t+G_{\sigma}(t)\mathrm{d}W(t),\quad
t\ge 0\label{20}
\end{equation}
where $\varepsilon(t)=(x^*(t),v^*(t))^T$, $W(t)=(W_1(t),W_2(t))^T$
and
$$
F_{\sigma}(t)=\left(\begin{array}{cc}
-h(t)(L_{\sigma(t)}+B_{\sigma(t)})&g(t)I_n\\-\gamma
h(t)(L_{\sigma(t)}+B_{\sigma(t)})&0
\end{array}\right),\qquad
G_{\sigma}(t)=h(t)\left(\begin{array}{c} G_{\sigma(t)}\\ \gamma
G_{\sigma(t)}
\end{array}\right):=h(t)\widetilde{G}_{\sigma(t)}.
$$

In the sequel, we show that under switching topology, the consensus
protocols (\ref{17}) and (\ref{18}) ensure that each follower tracks
the leader in mean square. We will use the following lemma.

\smallskip
\noindent\textbf{Lemma 2.}\cite{31}\itshape \quad Given $t\ge0$ and
suppose the digraph $\mathcal{G}_{\sigma(t)}$ is balanced. Then
$L_{\sigma(t)}+B_{\sigma(t)}+(L_{\sigma(t)}+B_{\sigma(t)})^T$ is
positive definite if and only if the vertex 0 is globally reachable
in $\overline{\mathcal{G}}_{\sigma(t)}$. \normalfont

\smallskip
The matrix
$L_{\sigma(t)}+B_{\sigma(t)}+(L_{\sigma(t)}+B_{\sigma(t)})^T$ plays
a key role in the convergence analysis below. Define
$\mu:=\min_{t\ge0}\big\{\lambda_{\min}(L_{\sigma(t)}+B_{\sigma(t)}+(L_{\sigma(t)}+B_{\sigma(t)})^T)\big\}$.

Prior to establishing the main result, we present a
condition analogous with Assumption (A2) in Section 3.1:\\
\textbf{(A5)}\quad There is a $\delta>0$, such that
$\frac{h(t)}{g(t)}>\frac1{2\gamma(1-\gamma^2)\mu}+\delta$ for $t\ge
0$.

\noindent\textbf{Remark 8.}\itshape \quad It is easily shown that
$\mu>0$ under the assumptions of Theorem 2 below by exploiting Lemma
2 and the fact that $\mathcal{H}$ is a finite set. This validates
the expression in Assumption (A5). \normalfont

\smallskip
\noindent\textbf{Theorem 2.}\itshape \quad For system (\ref{1}) with
the consensus protocols (\ref{17}) and (\ref{18}), if for any
$t\ge0$, $\mathcal{G}_{\sigma(t)}$ is balanced, and vertex 0 is
globally reachable in $\overline{\mathcal{G}}_{\sigma(t)}$, then
under Assumptions (A3)-(A5), we have
\begin{equation}
\lim_{t\rightarrow\infty}E\|\varepsilon(t)\|^2=0.\label{21}
\end{equation}
\normalfont

\smallskip
\noindent\textbf{Proof.} Let
$\widetilde{I}:=\left(\begin{array}{cc}I_n&-\gamma I_n\\
-\gamma I_n&I_n\end{array} \right)$. Obviously, we have
$\lambda_{\min}(\widetilde{I})=1-\gamma$ and
$\lambda_{\max}(\widetilde{I})=1+\gamma$. Hence $\widetilde{I}$ is a
positive definite matrix by recalling $0<\gamma<1$. Define a
Lyapunov function
$V(t)=\varepsilon^T(t)\widetilde{I}\varepsilon(t)$.

By It\^o formula and (\ref{20}), we have
$$
\mathrm{d}V(t)=\varepsilon^T(t)\big(\widetilde{I}F_{\sigma}(t)+F_{\sigma}^T(t)\widetilde{I}\big)\varepsilon(t)\mathrm{d}t+\mathrm{tr}(G_{\sigma}^T(t)\widetilde{I}G_{\sigma}(t))\mathrm{d}t+2\varepsilon^T(t)\widetilde{I}G_{\sigma}(t)\mathrm{d}W(t).
$$
Straightforward calculation yields
\begin{eqnarray*}
\widetilde{I}F_{\sigma}(t)+F_{\sigma}^T(t)\widetilde{I}&=&-h(t)\left(\begin{array}{cc}(1-\gamma^2)\big(L_{\sigma(t)}+B_{\sigma(t)}+(L_{\sigma(t)}+B_{\sigma(t)})^T\big)&-I_n\frac{g(t)}{h(t)}\\
-I_n\frac{g(t)}{h(t)}&2\gamma I_n\frac{g(t)}{h(t)}
\end{array}\right)\\
&:=&-h(t)\widetilde{Q}_{\sigma}(t).
\end{eqnarray*}
By using the Haynsworth inertia additivity formula \cite{30},
Assumption (A5), we get that $\widetilde{Q}_{\sigma}(t)$ is positive
definite with
$\nu:=\min_{t\ge0}\big\{\lambda_{\min}(\widetilde{Q}_{\sigma}(t))\big\}>0$.

Therefore, we have
\begin{equation}
\mathrm{d}V(t)\le-h(t)\frac{\nu}{1+\gamma}V(t)\mathrm{d}t+h^2(t)\mathrm{tr}\big(\widetilde{G}_{\sigma(t)}^T\widetilde{I}\widetilde{G}_{\sigma(t)}\big)\mathrm{d}t+2h(t)\varepsilon^T(t)\widetilde{I}\widetilde{G}_{\sigma(t)}\mathrm{d}W(t).\label{22}
\end{equation}

The remaining proofs are similar with those in Theorem 1 by noting
that $\mathcal{H}$ is a finite set, and hence omitted. $\Box$

\noindent\textbf{Remark 9.}\itshape \quad From Theorem 2 we see that
the designed protocol may guarantee the state of each follower
tracks that of the leader in mean square even under the switching
topology. \normalfont

\bigskip
\noindent{\Large\textbf{4. Numerical example}}
\smallskip

In this section, we provide a numerical simulation to illustrate the
theoretical results.

We consider a network consisting of four agents including one leader
labeled by vertex 0, as shown in Fig. 1. The digraph in this figure
is assumed to have $0-1$ weights. With simple calculation, it is not
hard to solve $P$ from Equation (\ref{13}) and obtain
$\lambda_{\max}(P)=0.9447$. We take $\sigma_{ij}=0.1$ for all $i,j$,
$h(t)=\frac1{t+2}$, $g(t)=\frac1{6(t+2)}$ and $\gamma=0.5$.
Therefore, Assumptions (A1)-(A4) in Theorem 1 hold.

The simulation results for the consensus errors $x^*$ and $v^*$ are
shown in Fig. 2 and Fig. 3 respectively, with initial value
$\varepsilon(0)=(-2,1.5,3,2,-1.5,-1)^T$. From Fig. 2 and Fig. 3, we
can see that three followers can eventually follow the leader.

\bigskip
\noindent{\Large\textbf{5. Conclusion}}
\smallskip

This paper studies a leader-following coordination problem for
multi-agent systems with a time-varying leader under measurement
noises. Although the state of the leader keeps changing and the
measured information by each follower is corrupted by white noises,
we propose a neighborhood-based protocol for each agent to follow
the leader. We present sufficient conditions for each follower to
track the leader in mean square under directed fixed topologies.
Sufficient conditions are also provided when the interaction
topology is switching and the subgraph formed by the followers is
balanced. Finally, numerical simulations are presented to illustrate
the theoretical results. Topics worth investigating in the future
include time-delay cases and the design of almost sure consensus
protocols.

\bigskip
\bigskip

\begin{center}
\textbf{Figure captions}

\end{center}

Fig. 1\quad Directed network $\overline{\mathcal{G}}$ of four agents
involving one leader. $\overline{\mathcal{G}}$ has $0-1$ weights.

Fig. 2\quad Consensus error $x^*$ for the agents.

Fig. 3\quad Consensus error $v^*$ for the agents.

\begin{center}
\setlength{\unitlength}{1mm}
\begin{picture}(60,60)
\put(2,58){0}\put(2,2){1}\put(58,2){3}\put(58,58){2}
\put(2,57){\vector(0,-1){51}}\put(5,59){\vector(1,0){52}}\put(5,5){\vector(1,1){52}}\put(55,55){\vector(-1,-1){50}}\put(5,3){\vector(1,0){52}}\put(59,6){\vector(0,1){51}}
\end{picture}
\end{center}

\centering \scalebox{0.5}{\includegraphics{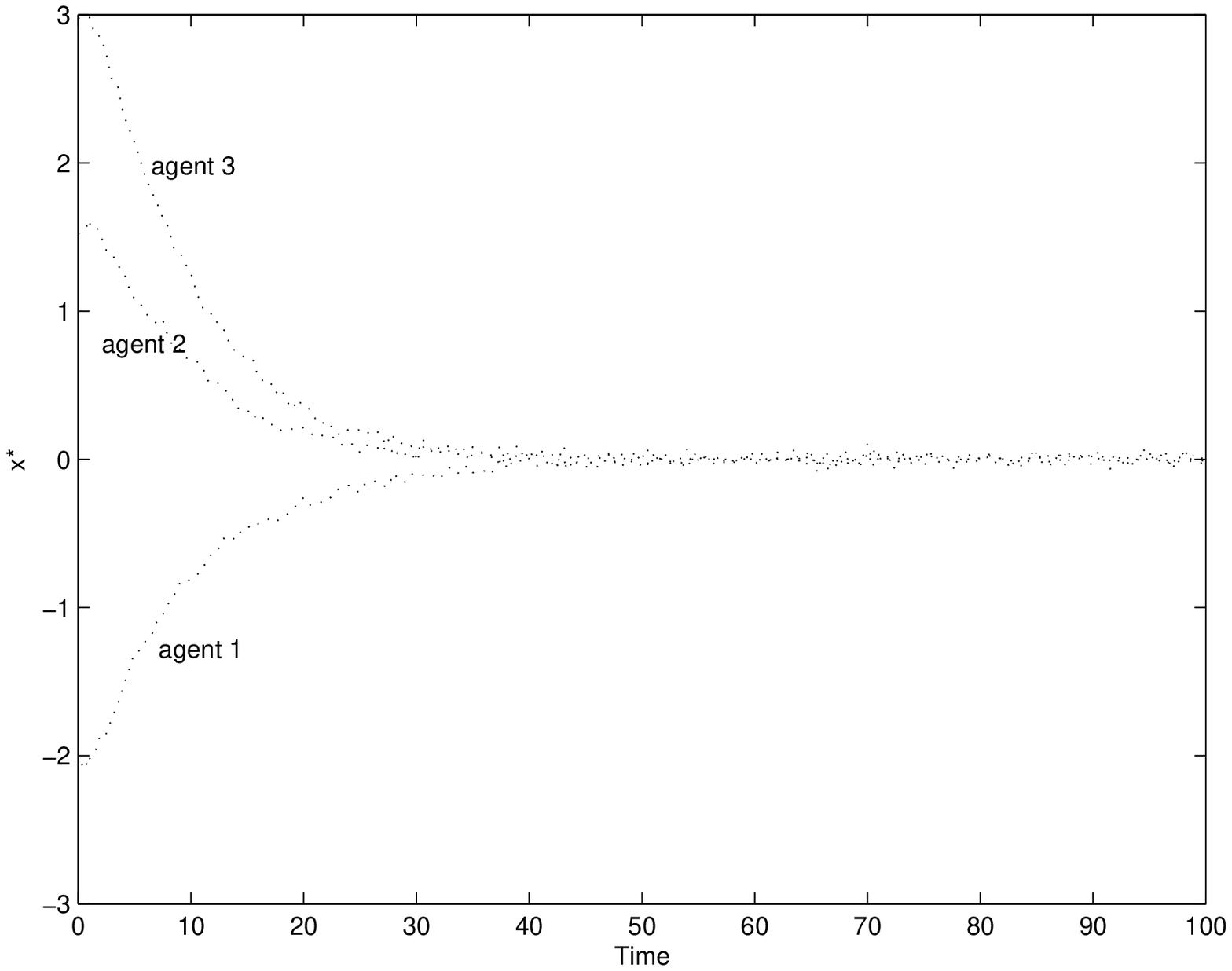}}
\label{fig_sim}

\centering \scalebox{0.5}{\includegraphics{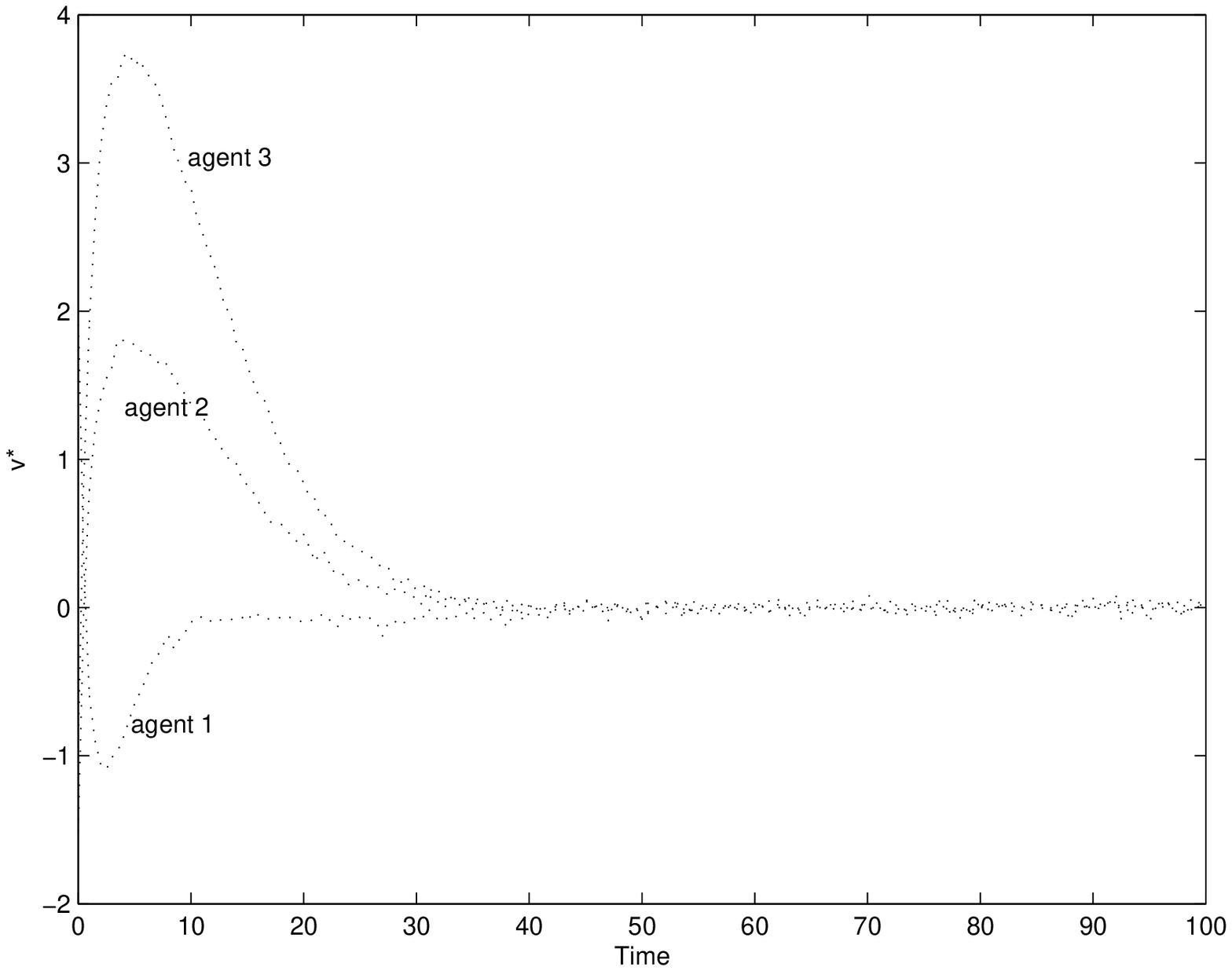}}
\label{fig_sim}

\end{document}